\documentstyle[11pt,epsf]{article}
\newcommand{\be}{\begin{equation}}
\newcommand{\ee}{\end{equation}}
\newcommand{\ben}{\begin{eqnarray}}
\newcommand{\een}{\end{eqnarray}}
\newcommand{\benn}{\begin{eqnarray*}}
\newcommand{\eenn}{\end{eqnarray*}}
\newcommand{\gb}{\bar{g}}

\newcommand{\mn}{{\mu\nu}}
\newcommand{\rs}{{\rho\sigma}}
\newcommand{\bD}{\mbox{\boldmath $\delta$}}
\begin{document}
\begin{titlepage}
\vspace{-20mm}
\begin{flushright}
KUCP-0128
\end{flushright}
\vspace{20mm}
\centerline{\LARGE Non-Trivial Ultraviolet Fixed Point in Quantum Gravity}
\vspace{5mm}
\centerline{{\large Wataru Souma}\footnote{\tt souma@phys.h.kyoto-u.ac.jp}}
\vspace{5mm}
\centerline{\large\it Faculty of Integrated Human Studies}
\vspace{0mm}
\centerline{\large\it Department of Physics, Kyoto University,
Kyoto 606-8501, Japan}
\vspace{50mm}
\begin{abstract}
The non-trivial ultraviolet fixed point in quantum gravity
is calculated by means of the exact renormalization group equation
in $d$-dimensions $(2\simeq d\leq4)$.
It is shown that the ultraviolet non-Gaussian fixed point which is
expected from the perturbativelly $\epsilon$-expanded calculations
in $2+\epsilon$ gravity theory remains in $d=4$.
Hence it is possible that quantum gravity is an asymptotically safe
theory and renormalizable in $2<d$. 
\end{abstract}
\end{titlepage}
\section{Introduction}
The successes of the perturbative renormalization group equation
have allowed for the progress in elementary particle physics.
However, there are many phenomena which have not been studied by this
perturbative method.
As is well known, quantum gravity (QG) must be treated non-perturbativelly.
To make QG a perturbativelly consistent theory
it is necessary to treat the infinite number of couplings.
This is because, in $d=4$,
the $L$-loop perturbative calculations in
Einstein gravity
cause divergences that are proportional
to the $L+1$ powers of the curvature tensor.
Thus to remove these divergences
infinite number of couplings are necessary.
However, if QG is an asymptotically safe theory,
it becomes renormalizable.\cite{Weinberg} \ An asymptotically safe
theory has ultraviolet (UV) non-Gaussian
fixed points (NGFP) on the hypersurface which separates the phase space.
Near that FP, $\infty-1,2,\cdots$ couplings become irrelevant.
Hence the renormalization group (RG) flows near the hypersurface are governed
by these irrelevant couplings and go toward the NGFP.
On the other hand, if the RG flows move away from the NGFP on the hypersurface,
the finite number of relevant couplings govern the theories.
In addition, if QG has an infrared (IR) Gaussian fixed point (GFP),
the RG flows move toward it.
As a result, the IR effective theories are described by the finite number of
couplings and become renormalizable.
Hence, whether or not QG is renormalizable depends on the existence
of the UV NGFP.

There are studies that suggest the existence of a UV NGFP.
One of these discussions is $2+\epsilon$ gravity theory which
applies the $\epsilon$-expanded perturbative calculation to QG in
$d=2+\epsilon.$\cite{KN01,KN02} \ The result of this theory suggests
that there exists a UV NGFP for
the Newton constant in $O(\epsilon)$, and it is expected that
it will remain in $d=4$.
This implies that QG is an asymptotically safe
theory and renormalizable for $d>2$.
However, the $\epsilon$-expansion is an asymptotic expansion. Hence
the large order behavior of $\epsilon$ is not reliable.
On the other hand, simplicial gravity
suggests that QG exhibits a first order phase transition in $d=3,4$.
This means that QG is not an
asymptotically safe theory or renormalizable in $d=3,4$.
Hence the results of $2+\epsilon$ gravity theory and simplicial gravity
are contradictory.

If this is actual, non-trivial phenomena must occur
in  the range $2<d\leq3$.
However the $\epsilon$-expansion and the
lattice calculation are not applicable in such dimensions.
Thus to solve this problem, other non-perturbative methods are needed.
The exact renormalization group equation (ERGE) \cite{Wilson}
is one non-perturbative method and does
not suffer from the constraints of the dimension,
if it is formulated in arbitrary dimensions.\cite{My} \ Recently, M. Reuter
formulated an ERGE for QG
in arbitrary dimensions.\cite{Reuter} \ Though the advantage that this
formulation is applicable in
any dimension has not been used in the investigation of the destiny of the
UV NGFP in $2+\epsilon$ gravity theory.
Hence, the purpose of this article is
to clarify, by means of Reuter's formulation, the possibility
that QG possesses a UV NGFP in $2<d\leq 4$.

This paper is organized as follows. In the next section the formulation
of the ERGE for QG is reviewed (for details see Ref.~\cite{Reuter}).
There, the $\beta$-functions for the Newton constant
and the cosmological constant are derived.
In \S3 the existence of the UV NGFP
and the perturbative limits of the ERGE are discussed using these
$\beta$-functions.
Section 4 is devoted to summary and discussion.

\section{Formulation and approximation of the ERGE}
\subsection{Quantization of gravity in the background field method}
In this subsection the Lagrangian ${\cal L}_{\rm GR}$
which is the general functional of the metric is quantized in 
the background field method.
The metric $\gamma_\mn$ is decomposed
as $\gamma_\mn=\gb_\mn+h_\mn$.
Here $\gb_\mn$ is the background field and is invariant under the BRST
transformations: $\bD_B\gb_\mn=0$.
Fluctuations around the background are denoted by $h_\mn$.
The BRST transformations of $h_\mn$, the anti-ghost field $\bar{C}_\mu$,
the ghost field $C^\mu$ and the $B$-field $B_\mu$ are given by
\benn
&&\bD_Bh_\mn=\kappa^{-2}{\cal L}_C\gamma_\mn,\hskip 5mm
\bD_B\bar{C}_\mu=B_\mu,\\
&&\bD_BC^\mu=\kappa^{-2}C^\nu\partial_\nu C^\mu,\hskip 3mm
\bD_BB_\mu=0.
\eenn
Here $\kappa$ is expressed in terms of the bare Newton constant $\bar{G}$ as
$\kappa=(32\pi\bar{G})^{-1/2}$. In addition, ${\cal L}_C$ represents the 
Lie derivative with respect to $C^\mu$.
The gauge fixing function $f_\mu$ is defined by
\[
f_\mu=F_\mu+\frac{\alpha}{2}B_\mu,
\]
where $\alpha$ is the gauge parameter.
Here $F_\mu$ fixes the gauges to the harmonic gauge and is given by
\[
F_\mu=\sqrt{2}\kappa{\cal F}^{\alpha\beta}_\mu[\gb]h_{\alpha\beta},
\hskip 5mm
{\cal F}_\mu^{\alpha\beta}[\gb]=
\delta^\beta_\mu\bar{g}^{\alpha\gamma}\bar{D}_\gamma
-\frac{1}{2}\bar{g}^{\alpha\beta}\bar{D}_\mu.
\]
Here $\bar{D}_\mu$ is the covariant derivative which is a function of $\gb_\mn$.
The gauge fixing term ${\cal L}_{\rm GF}$ and
the Faddeev-Popov ghost term ${\cal L}_{\rm FP}$ are given by
${\cal L}_{\rm GF}+{\cal L}_{\rm FP}
=\bD_{\rm B}\left(\bar{C}_\mu f^\mu\right)$.
Now the generating functional for the Green function and
that for the connected Green function are given by
\be
Z[J,\beta,\tau;\gb]=\exp W[J,\beta,\tau;\gb]=\int{\cal D}\Phi\exp\left\{
-S[\Phi;\gb]
-S_{\rm ES}
\right\},\label{eq:Z02}
\ee
where $B$-field integral is completed.
Here, the shorthand notation
$\Phi=\{h_\mn,C^\mu,\bar{C}_\mu\}$ for the fields and
$J=\{t^\mn,\bar\sigma_\mu,\sigma^\mu\}$ for the external source fields
is introduced.
The bare action $S[\Phi;\gb]$ in Eq. (\ref{eq:Z02}) is
\ben
S[\Phi;\gb]&=&\int d^dx{\cal L}_{\rm GR}[\gb+h]
+\frac{1}{2\alpha}\int d^dx\,\sqrt{\gb}\gb^\mn F_\mu F_\nu
-\int d^dx\sqrt{\gb}\bar{C}_\mu{\cal M}^\mu_\nu C^\nu\nonumber\\
&=&
S_{\rm GR}[h;\gb]+S_{\rm GF}[h;\gb]+S_{\rm FP}[h,C,\bar{C};\gb].
\label{eq:S01}
\een
Here ${\cal M}^\mu_\nu$ is given by
\[
{\cal M}[\gamma,\gb]^\mu_\nu=\gb^{\mu\rho}\gb^{\sigma\lambda}
\bar{D}_\lambda\left(\gamma_{\rho\nu}D_\sigma+\gamma_{\sigma\nu}D_\rho\right)
-\gb^{\rho\sigma}\gb^{\mu\lambda}\bar{D}_\lambda\gamma_{\sigma\nu}D_\rho.
\]
The last term in Eq. (\ref{eq:Z02}) is the external source term,
\be
S_{\rm ES}=
-\int d^dx\sqrt{\gb}\left\{
J\Phi
+\beta^\mn{\cal L}_C \gamma_\mn+\tau_\mu C^\nu\partial_\nu C^\mu
\right\},\label{eq:EXT}
\ee
where the external source fields
for the BRST transformations of $\Phi$ are included.

\subsection{Derivation of the exact renormalization group equation}
The scale-dependent generating functional for the connected Green function is
defined by
\be
\exp\{W_k[J,\beta,\tau;\gb]\}=\int{\cal D}\Phi\exp\{-S[\Phi;\gb]
-\Delta_kS[\Phi;\gb]-S_{\rm ES}\},
\label{eq:W}
\ee
where $k$ is the cutoff scale.
The cutoff action $\Delta_kS[\Phi;\gb]$ is given by
\ben
\Delta_kS[\Phi;\gb]&=&\frac{1}{2}\kappa^2\int d^dx\,\sqrt{\gb}\,h_\mn
(R_k^{\rm grav}[\gb])^{\mn\rho\sigma}h_{\rho\sigma}\nonumber\\
&&+\sqrt{2}\int d^dx\,\sqrt{\gb}\,\bar{C}_\mu
(R_k^{\rm gh}[\gb])C^\mu.
\een
Here the cutoff operators $R_k^{\rm grav}[\gb]$ and $R_k^{\rm gh}[\gb]$
are expressed as
\benn
R^{\rm grav}_k[\gb]&=&{\cal Z}_k^{\rm grav} k^2R^{(0)}(-\bar{D}^2/k^2),
\hskip 7mm {\cal Z}_k^{\rm grav}=Z_k^{\rm grav}\gb^\mn\gb^\rs,\\
R^{\rm gh}_k[\gb]&=&Z_k^{\rm gh}k^2R^{(0)}(-\bar{D}^2/k^2),
\eenn
where $Z_k^{\rm grav}$ and $Z_k^{\rm gh}$ are the renormalization factor of 
$h_\mn$ and the ghost field, respectively.
The cutoff function $R^{(0)}(u)$ must satisfy two conditions:
\be
R^{(0)}(0)=1,\hskip 5mm R^{(0)}(u\rightarrow\infty)=0.
\label{eq:cccond}
\ee
In this article $R^{(0)}(u)=u/(\exp(u)-1)$ is used.
The cutoff action $\Delta_kS$ goes to zero as $k\rightarrow0$,
because the cutoff operators $R_k^{\rm grav,gh}[\gb]$ go to zero
in this limit.
Thus $W_k$ coincides with the usual generating functional for
the connected Green function in this limit.

If Eq. (\ref{eq:W}) is differentiated with respect to $t=\ln k$, we obtain
\ben
-\partial_t W_k&=&
\frac{1}{2}\kappa^2\int d^dx\,\frac{1}{\sqrt{\gb}}
\frac{\delta W_k}{\delta t^\mn}(\partial_t R_k^{\rm grav})^{\mn\rho\sigma}
\frac{\delta W_k}{\delta t^{\rho\sigma}}\nonumber\\
&&+\frac{1}{2}\kappa^2\int d^dx\,\frac{1}{\sqrt{\gb}}
(\partial_t R^{\rm grav}_k)^{\mn\rho\sigma}
\frac{\delta^2 W_k}{\delta t^\mn\delta t^{\rho\sigma}}\nonumber\\
&&+\sqrt{2}\int d^dx\,\frac{1}{\sqrt{\gb}}
\frac{\delta W_k}{\delta \sigma^\mu}(\partial_t R_k^{\rm gh})
\frac{\delta W_k}{\delta \bar{\sigma}^\mu}\nonumber\\
&&-\frac{1}{2}\sqrt{2}\int d^dx\,\frac{1}{\sqrt{\gb}}
(\partial_t R_k^{\rm gh})
\left\{\frac{\delta^2W_k}{\delta \bar{\sigma}^\mu\delta \sigma^\mu}
-\frac{\delta^2W_k}{\delta\sigma^\mu\delta\bar{\sigma}^\mu}
\right\}.\label{eq:PARW}
\een
The functional derivatives of $W_k$ are regarded as the vertices.
Hence the first and third terms on the RHS of Eq. (\ref{eq:PARW}) correspond
to the dumbbell (tree) diagrams.
In addition, the second and
fourth terms on there correspond to the loop diagrams.
Effectively, the loop diagrams contribute to the low energy physics.
The introduction of the effective averaged action $\Gamma_k$ makes
Eq. (\ref{eq:PARW}) 1PI form.\cite{Wetterich} \ The effective averaged
action is defined by
\be
\Gamma_k[\varphi,\beta,\tau;\gb]=\int d^dx\,\sqrt{\gb}J(x)\varphi(x)
-W_k[J,\beta,\tau;\gb]-\Delta_kS[\varphi;\gb],\label{eq:EEAA}
\ee
where the shorthand notation $\varphi=\{\bar{h}_\mn,\xi^\mu,\bar\xi_\mu\}$
is used for the classical field $\varphi$.
The components of $\varphi$ are given
by
\[
\bar{h}_\mn=\frac{1}{\sqrt{\gb}}\frac{\delta W_k}{\delta t^\mn},\hskip 5mm
\xi^\mu=\frac{1}{\sqrt{\gb}}\frac{\delta W_k}{\delta\bar\sigma_\mu},\hskip 5mm
\bar\xi^\mu=\frac{1}{\sqrt{\gb}}\frac{\delta W_k}{\delta\sigma^\mu}.
\]
In addition, the classical field $g_\mn$ corresponding to $\gamma_\mn$ is
introduced by
\[
g_\mn(x)=\gb_\mn(x)+\bar{h}_\mn(x).
\]
If Eq. (\ref{eq:EEAA}) is inserted into Eq. (\ref{eq:PARW}), we have
\ben
\partial_t\Gamma_k&=&\frac{1}{2}{\rm Tr}\left[
\left(\Gamma_k^{(2)}+\kappa^2R_k^{\rm grav}\right)^{-1}_{\bar{h}\bar{h}}
(\partial_t\kappa^2R_k^{\rm grav})^{\mn\rho\sigma}\right]\nonumber\\
&&-\frac{1}{2}{\rm Tr}\left[\left\{
\left(\Gamma_k^{(2)}\!+\!\sqrt{2}R_k^{\rm gh}\right)^{-1}_{\bar{\xi}\xi}
\!\!-\left(\Gamma_k^{(2)}\!+\!\sqrt{2}R_k^{\rm gh}\right)^{-1}_{\xi\bar{\xi}}
\right\}(\partial_t\sqrt{2}R_k^{\rm gh})
\right].\label{eq:FE01}
\een
This is the ERGE for the effective averaged action.
Here $\Gamma_k^{(2)}$ is the Hessian
of $\Gamma_k$ with respect to the subscript.
For the same reason as that with $W_k$, $\Gamma_k$ coincides with the usual 1PI
effective action as $k\rightarrow0$.
In addition, at the cutoff scale $\Lambda$, which is the cutoff of the theory,
$\Gamma_k$ satisfies
\ben
\Gamma_\Lambda[\bar{h},\xi,\bar\xi;\beta,\tau;\gb]
&=&S_{\rm GR}[\bar{h}+\gb]+S_{\rm GF}[\bar{h};\gb]
+S_{\rm FP}[\bar{h},\xi,\bar\xi;\gb]
\nonumber\\
&&-\int d^dx\,\sqrt{\gb}\left\{\beta^\mn{\cal L}_\xi g_\mn
+\tau_\mu\xi^\nu\partial_\nu\xi^\mu\right\}.\label{eq:BC}
\een
This is the boundary condition to solve Eq. (\ref{eq:FE01})
(for details see Ref.~\cite{Reuter}).

\subsection{Constraints of the functional space}
To this point the breaking of the BRST invariance by the introduction of the
cutoff action has not been mentioned.
In the usual field theories, the Ward-Takahashi identity is
\[
0=\langle\bD_BS_{\rm ES}\rangle.
\]
However, the introduction of $\Delta_kS$ modifies this to
\[
0=\langle\bD_B S_{\rm ES}+\bD_B\Delta_kS\rangle.
\]
If this is written in terms of the effective averaged action, we obtain
\be
\int d^dx\frac{1}{\sqrt{\gb}}\left\{
\frac{\delta\Gamma'_k}{\delta\bar{h}_\mn}
\frac{\delta\Gamma'_k}{\delta\beta^\mn}
+\frac{\delta\Gamma'_k}{\delta\xi^\mu}
\frac{\delta\Gamma'_k}{\delta\tau_\mu}
\right\}
=Y_k\left(R_k^{\rm grav}[\gb],R_k^{\rm gh}[\gb]\right)
\label{eq:mwti},
\ee
where $\Gamma'_k=\Gamma_k-S_{\rm GF}$ has been introduced.
In the usual field theories, the RHS of Eq. (\ref{eq:mwti}) disappears.
Though the existence of the cutoff action makes it
proportional to $R_k^{\rm grav,gh}[\gb]$.
However, $Y_k$ goes to zero as $k\rightarrow0$, because
$R_k^{\rm grav,gh}[\gb]$ goes to zero in this limit.
Hence the usual field theories are recovered.
On the other hand, $Y_k$ does not disappear in the intermediate scale $k$.
Thus the BRST symmetry is broken in these scales.

Now to obtain the BRST invariant RG flows approximately,
the space of the action functionals is
truncated. As a first step towards such a truncation the evolution of the
ghost action is neglected.
Under this approximation the ansatz of the effective averaged
action is
\ben
\Gamma_k[g,\gb,\xi,\bar\xi;\beta,\tau]
&=&\bar\Gamma_k[g]+\hat\Gamma_k[g,\gb]+S_{\rm GF}[g-\gb;\gb]
+S_{\rm FP}[g-\gb,\xi,\bar\xi;\gb]\nonumber\\
&&-\int d^dx\,\sqrt{\gb}\left\{\beta^\mn{\cal L}_\xi g_\mn
+\tau_\mu\xi^\nu\partial_\nu\xi^\mu\right\}.\label{eq:exgamma}
\een
Here $S_{\rm GF}$ and $S_{\rm FP}$ are in the same form as in the bare action.
The coupling to the BRST variations also has the same form
as in the bare action.
The remaining term is decomposed into $\bar\Gamma_k[g]$ and $\hat\Gamma_k[g,\gb]$.
The latter term $\hat\Gamma_k$ contains
the deviations for $g_\mn\neq\gb_\mn$. Therefore
$\hat\Gamma[g,\gb=g]=0$, by definition. 
This term is interpreted as the quantum corrections to $S_{\rm GF}$.
Equation (\ref{eq:exgamma}) satisfies the boundary condition
Eq. (\ref{eq:BC}) at $k=\Lambda$
if these terms satisfy
\[
\bar\Gamma_\Lambda=S_{\rm GR},\hskip 5mm
\hat\Gamma_\Lambda=0.
\]
This condition suggests that setting $\hat\Gamma_k=0$ for all $k$ is
a candidate for the first approximation. 
If Eq. (\ref{eq:exgamma}) is inserted into Eq. (\ref{eq:mwti}), we have
\be
\int d^dx{\cal L}_\xi g_\mn\frac{\delta\hat\Gamma_k[g,\gb]}{\delta g_\mn(x)}
=-Y_k\left(R_k^{\rm grav}[\gb],R_k^{\rm gh}[\gb]\right). \label{eq:MODWT}
\ee
The RHS of  Eq. (\ref{eq:MODWT}) is regarded as the higher loop corrections
(for details see Ref.~\cite{Reuter}).
Hence to neglect $Y_k$ is acceptable in the first approximation.
This is consistent with setting $\hat\Gamma_k=0$ for all scales.
These approximation means that the RG flows are projected on the background
spaces.
On the background spaces the BRST invariance is recovered.
As a result, the projected RG flows are the BRST invariant.

If Eq. (\ref{eq:exgamma}) is inserted into Eq. (\ref{eq:FE01}), we find
\ben
\partial_t\Gamma_k[g,\gb]
&=&\frac{1}{2}{\rm Tr}\left[
\left(\kappa^{-2}\Gamma_k^{(2)}[g,\gb]+R_k^{\rm grav}[\gb]\right)^{-1}
(\partial_t R_k^{\rm grav}[\gb])\right]\nonumber\\
&&-{\rm Tr}\left[
\left(-{\cal M}[g,\gb]+R_k^{\rm gh}[\gb]\right)^{-1}
(\partial_tR_k^{\rm gh}[\gb])
\right].\label{eq:FE02}
\een
This is the truncated ERGE for $\Gamma_k[g,\gb]$.
Here $\Gamma_k[g,\gb]$ is given by
\ben
\Gamma_k[g,\gb]&=&\Gamma_k[g,\gb,0,0;0,0]\nonumber\\
&=&\bar\Gamma_k[g]+S_{\rm GF}[g-\gb;\gb]+\hat\Gamma[g,\gb].
\label{eq:prog}
\een
In addition, $\Gamma_k^{(2)}$ is the Hessian of $\Gamma_k[g,\gb]$ with 
respect to $g_\mn$ at fixed $\gb_\mn$.

\subsection{Einstein-Hilbert truncation}
To actually solve Eq. (\ref{eq:FE02}) is very
difficult. Hence to make the problem easier, the effective averaged action
must be approximated. The most naive approximation
is to expand the effective averaged action in terms of the operators which
are invariant under general coordinate transformations.
The first step of this approximation is to take $S_{\rm GR}$ as
the Einstein-Hilbert action:
\[
S_{\rm GR}=\frac{1}{16\pi\bar{G}}\int d^dx\,\sqrt{g}\left\{
-R(g)+2\bar{\lambda}\right\},
\]
where $\bar\lambda$ is the bare cosmological
constant. The scale dependent couplings are defined by
\[
\bar{G}\rightarrow G_k=Z_{Nk}^{-1}\bar{G},
\hskip 5mm \bar\lambda\rightarrow\bar\lambda_k,
\hskip 5mm \alpha\rightarrow\alpha_k=Z_{NK}^{-1}\alpha.
\]
In this article $\alpha=1$ is used. From Eq. (\ref{eq:prog}),
the effective averaged action becomes
\ben
\Gamma_k[g,\gb]&=&2\kappa^2Z_{Nk}\int d^dx\,\sqrt{g}
\left\{-R(g)+2\bar\lambda_k\right\}\nonumber\\
&&+\kappa^2Z_{Nk}\int d^dx\,\sqrt{\gb}\,\gb^\mn
({\cal F}^{\alpha\beta}_\mu g_{\alpha\beta}) 
({\cal F}^{\rho\sigma}_\nu g_{\rho\sigma}).\label{eq:EAA}
\een
Here $\hat\Gamma_k$ is neglected.
If Eq. (\ref{eq:EAA}) is differentiated with respect to $t$
and projected at $g_\mn=\gb_\mn$, we have
\be
\partial_t\Gamma_k[g,g]=2\kappa^2\int d^dx\,\sqrt{g}\left[
-R(g)\partial_tZ_{Nk}+2\partial_t(Z_{Nk}\bar\lambda_k)\right].
\label{eq:LHS}
\ee
This is the LHS of Eq. (\ref{eq:FE02}).
The next step is to get the RHS of Eq. (\ref{eq:FE02}).
The effective averaged action is expanded in terms of $\bar{h}_\mn$.
For the quadratic term of $\bar{h}_\mn$, $\bar{h}_\mn$ is decomposed as
\[
\bar{h}_\mn=\hat{h}_\mn+\frac{1}{d}\gb_\mn\phi.
\]
Here $\hat{h}_\mn$ is the traceless part of $\bar{h}_\mn$: $\gb^\mn\hat{h}_\mn=0$.
On the other hand, $\phi$ is the trace part of $\bar{h}_\mn$:
$\phi=\gb^\mn\bar{h}_\mn$.
In addition, the function $({\cal Z}_k^{\rm grav})^{\mn\rho\sigma}$ in the
cutoff operator is decomposed as
\be
({\cal Z}_k^{\rm grav})^{\mn\rho\sigma}=
\left[(I-P_\phi)^{\mn\rho\sigma}-\frac{d-2}{2d}P_\phi^{\mn\rho\sigma}
\right]Z_{Nk},\label{eq:projz}
\ee
where
\[
P_\phi^{\mn\rho\sigma}=d^{-1}\gb^\mn\gb^{\rho\sigma}.
\]
The first term on the RHS of Eq. (\ref{eq:projz}) projects the traceless part.
The second term there projects the trace part.
To proceed with the calculations, the background is fixed by the maximally
symmetric space.
Hence the Riemann tensor $\bar{R}_{\mn\rho\sigma}$ and the
Ricchi tensor $\bar{R}_\mn$ are parameterized in terms of the $c$-number
scalar curvature $\bar{R}$. 
In this background, $\bar{R}$ is regarded as
not a function of the metric, but a number.
In the following the projection at $g_\mn=\gb_\mn$ is applied. Therefore,
the bars of the metric and curvature scalar are omitted.
From these manipulations, if the RHS of Eq. (\ref{eq:FE02}) is
denoted by ${\cal S}_k(R)$, we have
\ben
{\cal S}_k(R)
&=&{\rm Tr}_T\left[{\cal N}({\cal A}+C_TR)^{-1}\right]
+{\rm Tr}_S\left[{\cal N}({\cal A}+C_SR)^{-1}\right]\nonumber\\
&&-2{\rm Tr}_V\left[{\cal N}_0({\cal A}_0+C_VR)^{-1}\right],\label{eq:QUAD04}
\een
with
\[
C_T=\frac{d(d-3)+4}{d(d-1)},\hskip 5mm
C_S=\frac{d-4}{d},\hskip 5mm
C_V=-\frac{1}{d}.
\]
In Eq. (\ref{eq:QUAD04}), ${\cal A}$ and ${\cal N}$ are given by
\benn
{\cal A}&=&-D^2+k^2R^{(0)}(-D^2/k^2)-2\bar\lambda_k,\\
{\cal N}&=&\left(1-\eta(k)/2\right)k^2R^{(0)}(-D^2/k^2)
+D^2R^{(0)\prime}(-D^2/k^2),
\eenn
where the anomalous dimension $\eta(k)$ is defined by
$\eta(k)=-\partial_t\ln Z_{Nk}$.
In addition, ${\cal A}_0$ and ${\cal N}_0$ in Eq. (\ref{eq:QUAD04}) are
defined similarly to ${\cal A}$ and ${\cal N}$, except for $\bar\lambda_k=0$ and
$\eta(k)=0$. 

Now to obtain the coefficients of $\sqrt{g}$ and $\sqrt{g}R$, the RHS of
Eq. (\ref{eq:QUAD04}) is expanded in terms of the scalar curvature $R$:
\ben
{\cal S}_k(R)
&=&{\rm Tr}_T\left[{\cal N}{\cal A}^{-1}\right]
+{\rm Tr}_S\left[{\cal N}{\cal A}^{-1}\right]
-2{\rm Tr}_V\left[{\cal N}_0{\cal A}_0^{-1}\right]\nonumber\\
&&-R\left(C_T{\rm Tr}_T\left[{\cal N}{\cal A}^{-2}\right]
+C_S{\rm Tr}_S\left[{\cal N}{\cal A}^{-2}\right]\right.\nonumber\\
&&\left.-2C_V{\rm Tr}_V\left[{\cal N}_0{\cal A}_0^{-2}\right]
\right)+O(R^2).\label{eq:QUAD05}
\een
In addition to calculating the traces in Eq. (\ref{eq:QUAD05}),
the heat kernel expansion is applied.
Hence comparison with the results of these manipulations and
Eq. (\ref{eq:LHS}) gives the differential equations for
$\partial_tZ_{Nk}$ and $\partial_t(Z_{Nk}\bar\lambda_k)$.
If these equations are expressed in terms of the dimensionless couplings
of the Newton constant $g_k$ and the cosmological constant $\lambda_k$,
the $\beta$-functions for these are given by
\ben
\partial_t g_k=\beta_g&=&[d-2+\eta(k)]g_k,\label{eq:BETAG}\\
\partial_t\lambda_k=\beta_\lambda&=&-(2-\eta)\lambda_k
+\frac{1}{2}g_k(4\pi)^{1-d/2}\left\{2d(d+1)\Phi_{d/2}^1(-2\lambda_k)
\right.\nonumber\\
&&\left.-d(d+1)\eta(k)\widetilde\Phi_{d/2}^1(-2\lambda_k)-8d\Phi_{d/2}^1(0)
\right\}.\label{eq:BETAL}
\een
Here the dimensionless couplings are defined by
\[
g_k=k^{d-2}G_k=k^{d-2}Z_{Nk}^{-1}\bar{G},\hskip 5mm
\lambda_k=k^{-2}\bar\lambda_k.
\]
In this case, the anomalous dimension is expressed as
\be
\eta(k)=
\frac{g_kB_1(d,\lambda_k)}{1-g_kB_2(d,\lambda_k)}.\label{eq:eta}
\ee
The new functions $B_1(d,\lambda_k)$ and $B_2(d,\lambda_k)$ are defined by
\benn
B_1(d,\lambda_k)&=&\frac{1}{3}(4\pi)^{1-d/2}
\left[d(d+1)\Phi_{d/2-1}^1(-2\lambda_k)
-6d(d-1)\Phi_{d/2}^2(-2\lambda_k)\right.\nonumber\\
&&\left.-4d\Phi_{d/2-1}^1(0)-24\Phi_{d/2}^2(0)\right],\\
B_2(d,\lambda_k)&=&-\frac{1}{6}(4\pi)^{1-d/2}\left[
d(d+1)\widetilde\Phi_{d/2-1}^1(-2\lambda_k)\right.\nonumber\\
&&\left.-6d(d-1)\widetilde\Phi_{d/2}^2(-2\lambda_k)
\right].
\eenn
The functions $\Phi^p_n(w)$ and $\widetilde\Phi^p_n(w)$ are concerned with
the integrals of the cutoff function and given by
\benn
\Phi_n^p(w)&=&\frac{1}{\Gamma(n)}\int_0^\infty dz z^{n-1}
\frac{R^{(0)}(z)-zR^{(0)\prime}(z)}{[z+R^{(0)}(z)+w]^p},\\
\widetilde\Phi_n^p(w)&=&\frac{1}{\Gamma(n)}\int_0^\infty dz z^{n-1}
\frac{R^{(0)}(z)}{[z+R^{(0)}(z)+w]^p}.
\eenn
In addition, the relation
\be
\Phi_0^p(w)=\widetilde\Phi_0^p(w)
=\frac{1}{(1+w)^p}\label{eq:MP}
\ee
is fulfilled for any cutoff functions.

\section{Ultraviolet fixed points in $2\simeq d\leq4$}
\subsection{The case $\lambda_k=0$}
To see the behavior of the Newton constant and to
simplify the problem, the cosmological constant is ignored.
This approximation is applicable if the cosmological constant is
much smaller than the cutoff scale $k$.
In this case the phase structure is described by
the $\beta$-function for the Newton constant only:
\be
\partial_tg_k=\beta_g=\left[d-2+\eta^\prime(k)\right]g_k.\label{eq:L0BETA}
\ee
In contrast to Eq. (\ref{eq:eta}), the anomalous dimension
is denoted by $\eta^\prime(k)$ and given by
\be
\eta^\prime(k)=
\frac{g_kB_1(d,0)}{1-g_kB_2(d,0)}\label{eq:etap}.
\ee

Figure 1 (a) and (b) are numerically calculated results
of this $\beta$-function in some dimensions.
In Fig. 1 (a), the dotted line is the result in $d=1.9$.
This $\beta$-function has two FPs. 
One is the UV GFP, and the other is the IR NGFP, which exists in
a negative coupling region.
If only a positive coupling region is considered,
in $d=1.9$, QG has one phase and
becomes an asymptotically free theory.
The solid line in the figure is the result in $d=2$.
This $\beta$-function has one FP.
If the same coupling region is considered, QG has one
phase and becomes an asymptotically free theory in $d=2$.
The dashed line in the figure is the result in $d=2.1$.
This $\beta$-function has two FPs.
One is the IR GFP and the other is the UV NGFP.
In contrast to the result in $d=1.9$,
the NGFP exists in a positive coupling region.
Thus QG has two phases in this region.
One corresponds to the weak coupling phase, and the other corresponds to the
strong coupling phase. These results for the structure of the phase space
coincide with the ordinary perturbative calculation in $2+\epsilon$
gravity theory. 
This result suggests that this UV NGFP will remain in $d=4$.
However, the usual $\epsilon$-expansion does not predict
the possibility that UV NGFP in $d=2.1$ remains in higher dimensions.
On the other hand, Eq. (\ref{eq:L0BETA}) is free from a constraint of
the dimensionality.
Hence this $\beta$-function is applicable even if the dimension is
greater than 2.
Figure 1 (b) is the result in $2\leq d\leq4$.
This figure suggests that the UV NGFP in $d=2.1$ remains in $d=4$.
Hence it is expected that QG will have a UV NGFP in $2<d\leq4$.
This NGFP separates the positive coupling region into two phases.
One is the weak coupling phase, and the other is the strong one.

\begin{figure}[t]
\epsfxsize=0.49\textwidth
\begin{center}
\leavevmode
\epsffile{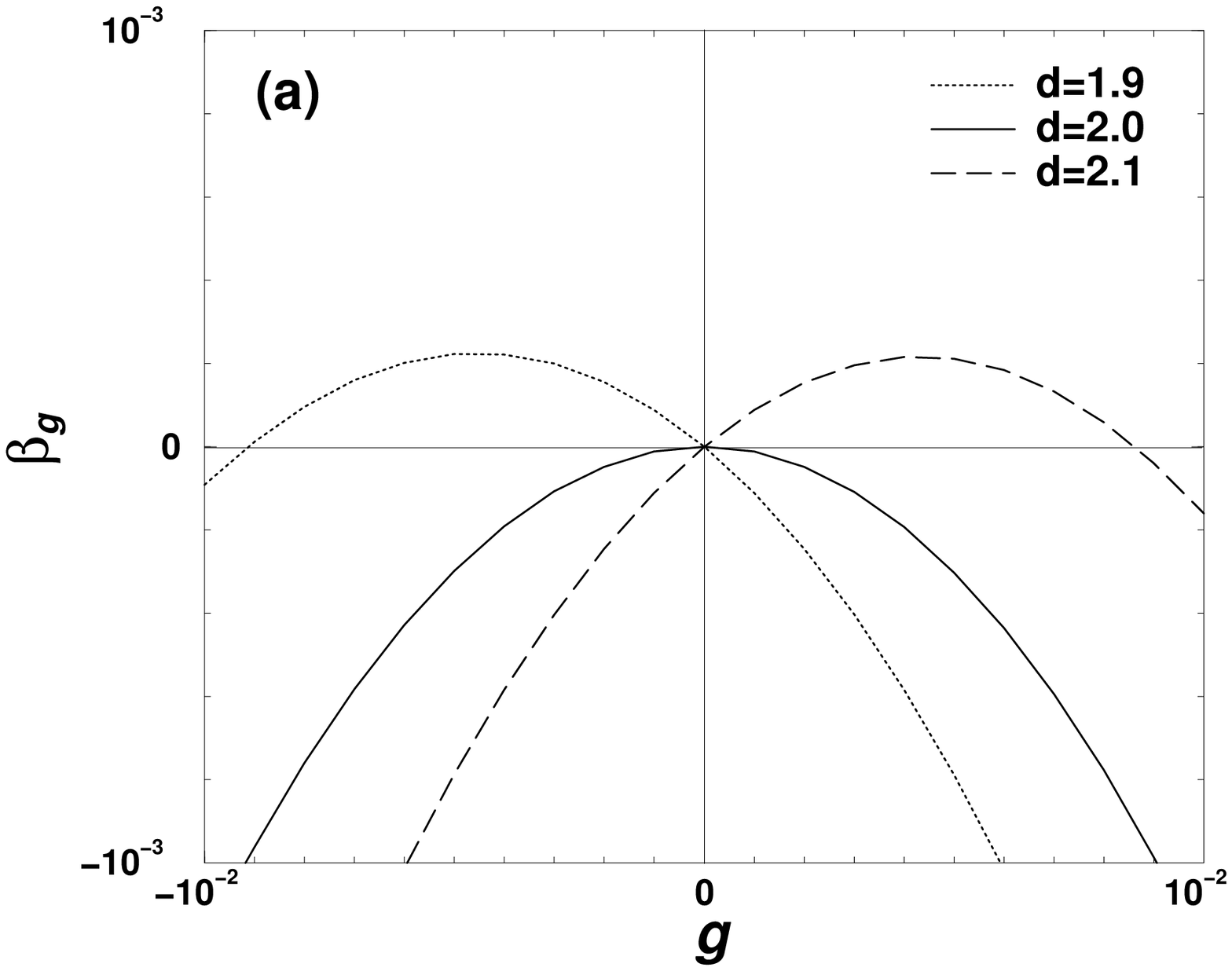}
\epsfxsize=0.48\textwidth
\epsffile{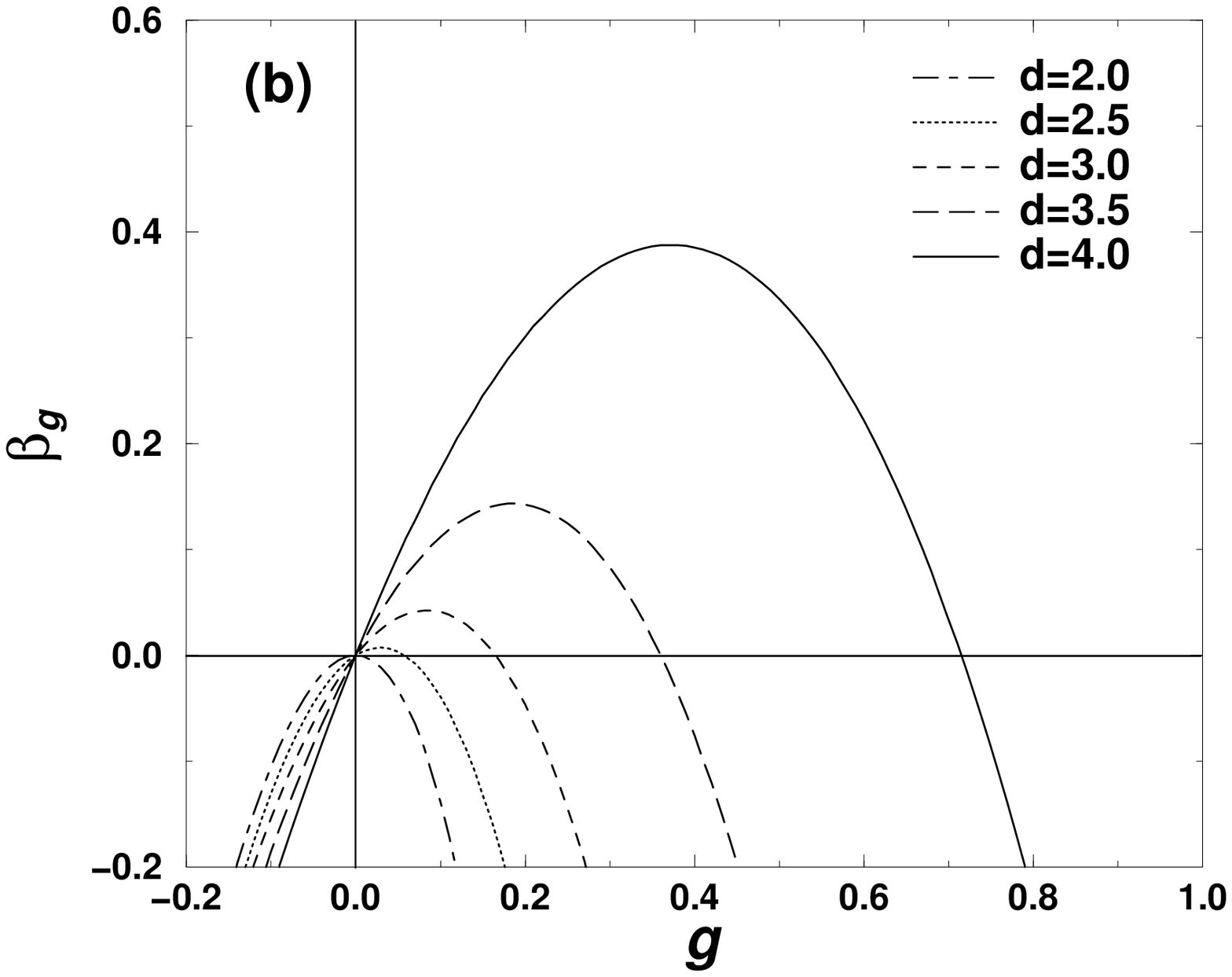}
\end{center}
\parbox{160mm}{\footnotesize
Fig. 1 : The numerically calculated $\beta$-functions for the Newton constant
in $d=1.9,2,2.1$ (a) and $d=2,2.5,3,3.5,4$ (b).}
\end{figure}

The numerical calculations of the RG flows in $d=4$ are shown in Fig. 2.
These were derived by numerically solving the differential equation
Eq. (\ref{eq:L0BETA}) under some initial conditions.
The horizontal axis is the scale $t$.
Small $t$ correspond to the IR region, and large $t$ correspond
to the UV region.
The horizontal lines are the UV NGFP and the IR GFP.
The dotted lines are the flows in
the negative coupling region.
The solid lines are the flows in the weak coupling phase.
These converge to the IR GFP.
The dashed lines are the flows in the strong
coupling phase.
These two phases are separated by the UV NGFP.
\begin{figure}[t]
\epsfxsize=8.5cm
\centerline{\epsfbox{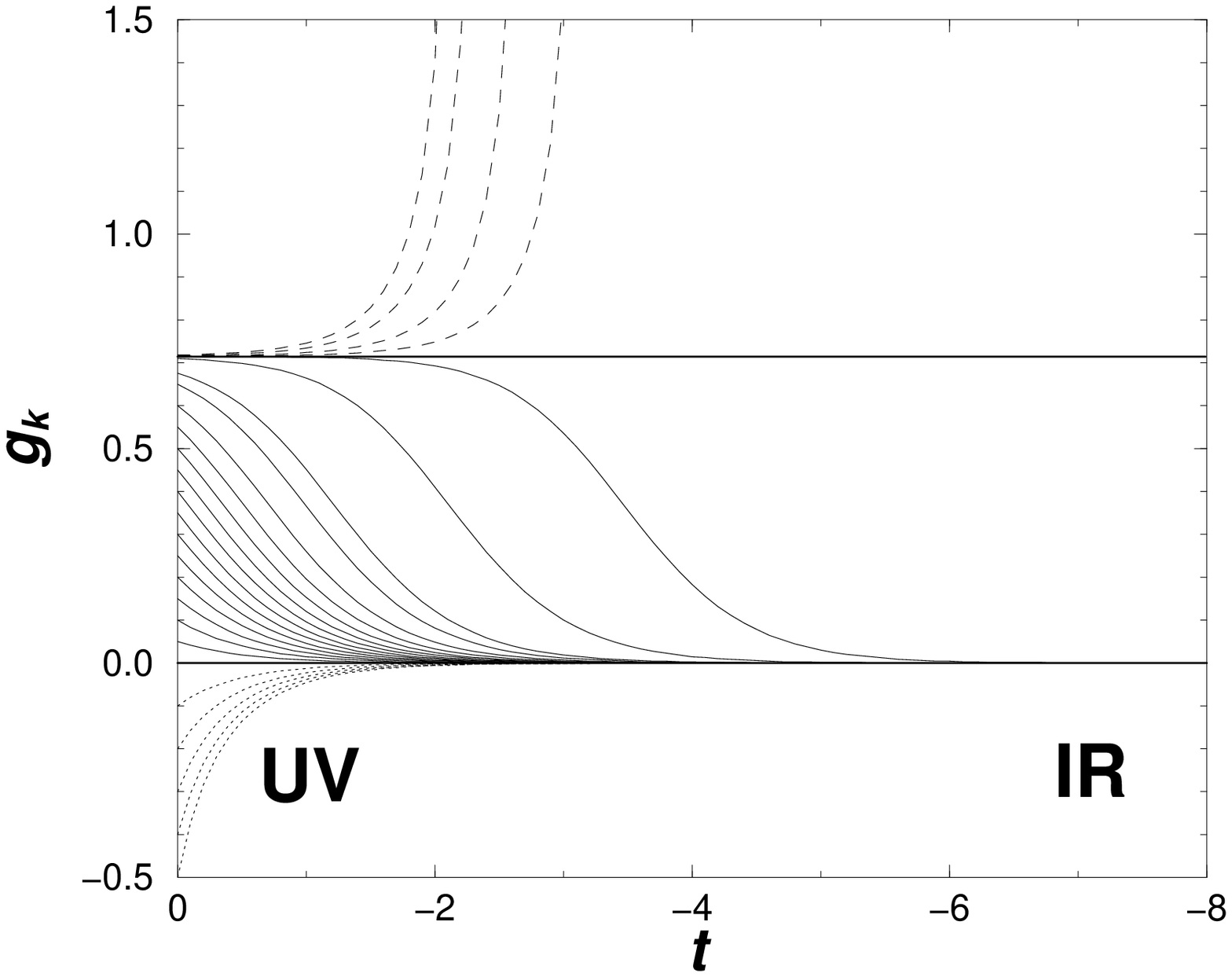}}
\parbox{160mm}{\footnotesize
Fig. 2 : The flows of the Newton constant $g_k$ in $d=4$.
The horizontal axis is the scale $t$.
Small $t$ correspond to the IR region, and large $t$ correspond
to the UV region.
The horizontal lines are the UV NGFP and the IR GFP.
The dotted lines are the flows in the negative coupling region.
The solid lines are the flows in the weak coupling phase.
The dashed lines are the flows in the strong coupling phase.
}
\end{figure}

The FPs $g^*$ satisfy
\[
0=\partial_tg^*=[d-2+\eta^\prime_*]g^*.
\]
Thus the solutions are $g^*=0$ and $\eta^\prime_*=2-d$. The former solution 
corresponds to the GFP and the latter is a candidate for the NGFP.
If Eq. (\ref{eq:etap}) is inserted into the latter condition, we obtain
\be
g^*=\frac{2-d}{B_1(d,0)+(2-d)B_2(d,0)}\label{eq:condfpg}.
\ee
This solution becomes zero in $d=2$. Therefore, GFP are degenerate
in this dimension.
On the other hand this solution is the NGFP in $d\neq2$.
The numerical calculation of Eq. (\ref{eq:condfpg}) in $2\leq d\leq4$
is shown as the solid line in Fig. 3.
The dashed line in this figure is the result of $2+\epsilon$ gravity
theory in the harmonic gauge up to $O(\epsilon)$.
These results coincide in $d\simeq2$.
However, the difference of the position for the UV NGFP
becomes large as $d\rightarrow4$.
To see the correspondence between the ERGE and $2+\epsilon$ gravity
theory, the perturbative limit of the ERGE is considered in $d=2+\epsilon$.
In this case, Eq. (\ref{eq:condfpg}) becomes
\[
g^*=\frac{-\epsilon}{B_1(2+\epsilon,0)-\epsilon B_2(2+\epsilon,0)}.
\]
If this equation is expanded in $\epsilon$, we have
\be
g^*\simeq-\frac{1}{B_1^{(0)}(2,0)}\epsilon+O(\epsilon^2).\label{eq:fpg2e}
\ee
Here $B_i(2+\epsilon,0),\;(i=1,2)$ is expanded as
\[
B_i(2+\epsilon,0)
=B_i^{(0)}(2,0)+B_i^{(1)}(2,0)\epsilon+\cdots,
\]
where
\[
B_1^{(0)}(2,0)=\frac{1}{3}\left[-2\Phi_0^1(0)-36\Phi_1^2(0)\right]=-\frac{38}{3}.
\]
This is because $\Phi_0^1(0)=1$, from Eq. (\ref{eq:MP}), and
$\Phi_1^2(0)=1$ for any cutoff functions.
Hence
\be
g^*\simeq\frac{3}{38}\epsilon+O(\epsilon^2).\label{eq:cc0gfp}
\ee
This coincides with the result of $2+\epsilon$ gravity theory in
the harmonic gauge up to $O(\epsilon)$.\cite{KN01} \ In addition, this
result is independent of the shape of the cutoff functions
up to this order.
\begin{figure}[t]
\epsfxsize=8.5cm
\centerline{\epsfbox{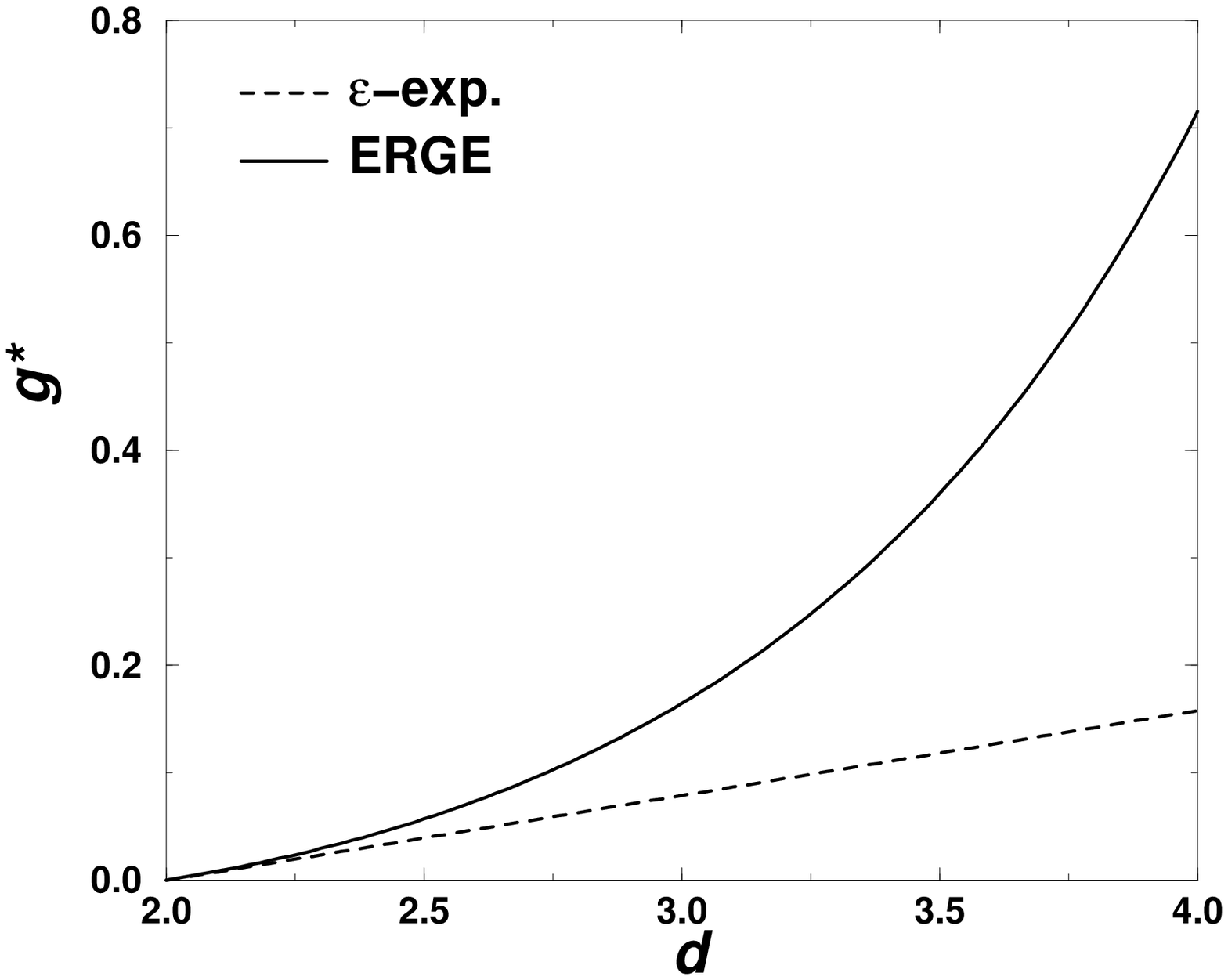}}
\parbox{160mm}{\footnotesize
Fig. 3 : The position of the UV NGFP for the Newton constant in $2\leq d\leq4$.
The solid line is the numerical calculation of Eq. (3$\cdot$3).
The dashed line is the result of $2+\epsilon$ gravity theory
in the harmonic gauge up to $O(\epsilon)$.
}
\end{figure}

\subsection{The case $\lambda_k\neq0$}
If the cosmological constant is considered,
a candidate for the NGFP becomes
\be
g^*=\frac{2-d}{B_1(d,\lambda^*)+(2-d)B_2(d,\lambda^*)},\label{eq:ngfpg}
\ee
where the FP of the cosmological constant is denoted by $\lambda^*$, which
must satisfy $\partial_t\lambda^*=0$.
If Eq. (\ref{eq:ngfpg}) and $\eta_*=2-d$ 
are inserted into Eq. (\ref{eq:BETAL}), we have
\ben
0=\beta_{\lambda^*}&=&-d\lambda^*+\frac{1}{2}
\frac{(2-d)(4\pi)^{1-d/2}}{B_1(d,\lambda^*)+(2-d)B_1(d,\lambda^*)}
\left\{2d(d+1)\Phi_{d/2}^1(-2\lambda^*)\right.\nonumber\\
&&\left.
-d(d+1)(2-d)\widetilde\Phi_{d/2}^1(-2\lambda^*)-8d\Phi_{d/2}^1(0)
\right\}.\label{eq:BETAL-A}
\een
This equation is a function
of $\lambda^*$ and $d$. Hence, if $\lambda^*$ is calculated
for each dimension,
the UV NGFP of the Newton constant can be derived from Eq. (\ref{eq:ngfpg}).
Hence the UV NGFP is expressed as $(g^*,\lambda^*)$.
The numerically calculated results for the UV NGFP $(g^*,\lambda^*)$ are
shown in Fig. 4.
In this figure the filled circles and the numbers beside these represent the
space-time dimensions.
From this figure it is concluded that
QG has a UV NGFP in $2<d\leq4$, even if the cosmological constant
is taken into account.
Hence it is possible that QG is an asymptotically safe theory
and renormalizable in $2<d\leq4$.

\begin{figure}[bht]
\epsfxsize=8.5cm
\centerline{\epsfbox{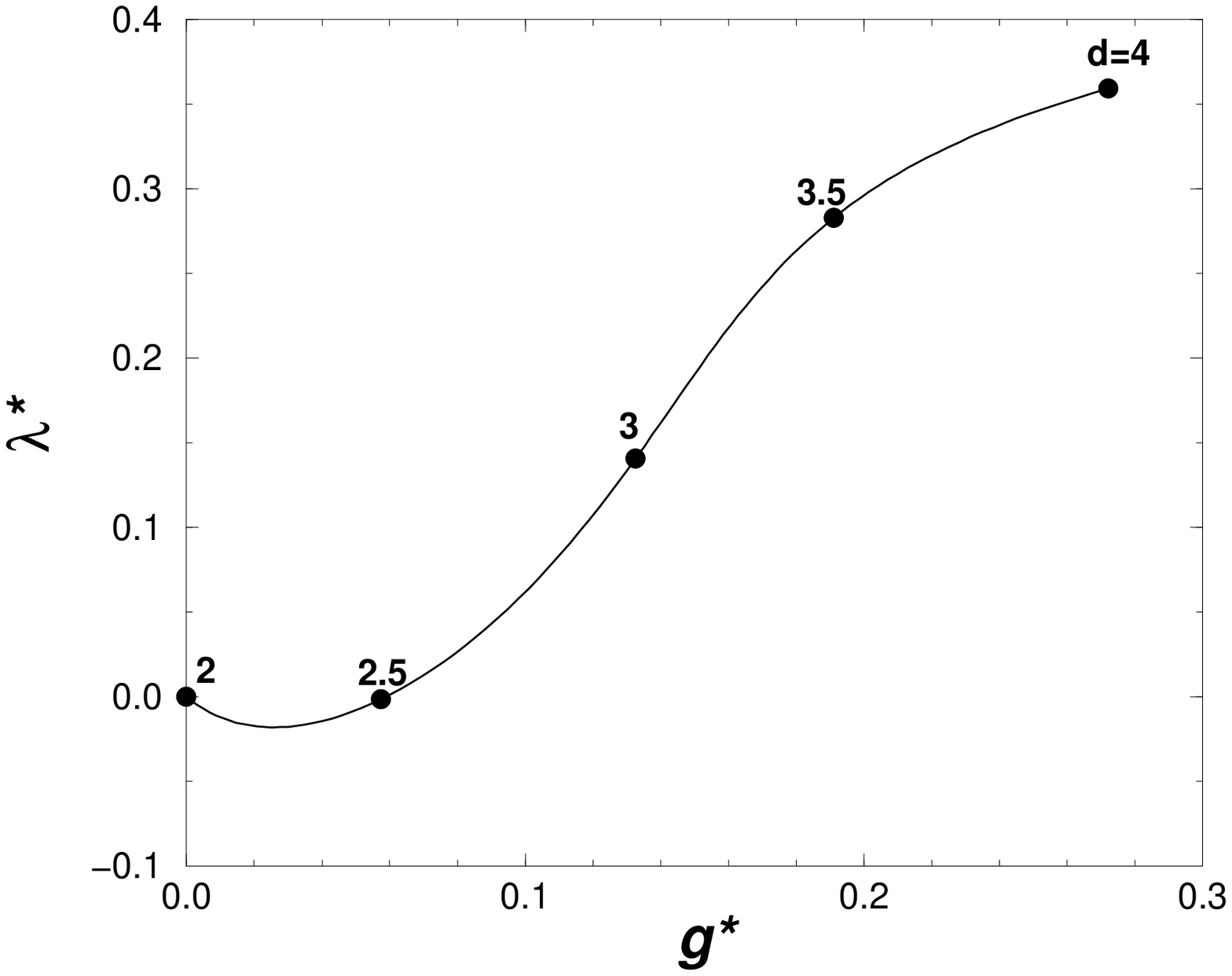}}
\parbox{160mm}{\footnotesize
Fig.4 : The numerical calculation of the UV NGFP $(g^*,\lambda^*)$
in $2\leq d\leq4$.
The filled circles and the numbers beside them indicate
the space-time dimension $d$.
}
\end{figure}

\newpage
In $d=2+\epsilon$, the UV NGFP $(g^*,\lambda^*)$ exists in $O(\epsilon)$.
Hence, Eq. (\ref{eq:ngfpg}) becomes
\be
g^*=\frac{-\epsilon}{B_1(2+\epsilon,\epsilon)
-\epsilon B_2(2+\epsilon,\epsilon)}.\label{eq:epsfpgc}
\ee
If Eq. (\ref{eq:epsfpgc}) is expanded in $\epsilon$, we obtain
\be
g^*\simeq-\frac{1}{B_1^{(0)}(2,0)}\epsilon+O(\epsilon^2).\label{eq:ccn0gfp}
\ee
As mentioned in the previous subsection, $B_i(2+\epsilon,\epsilon)$
is expanded in $\epsilon$:
$B_i(2+\epsilon,\epsilon)
\simeq B_i^{(0)}(2,0)+B_i^{(1)}(2,0)\epsilon+\cdots$.
Equation (\ref{eq:ccn0gfp}) coincides with
Eq. (\ref{eq:fpg2e}) and Eq. (\ref{eq:cc0gfp}).
Thus the UV NGFP for the Newton constant coincides with the
result of $2+\epsilon$ gravity theory in the harmonic gauge
up to $O(\epsilon)$.
In addition this result is independent of the shape of the cutoff functions.
The UV NGFP for the cosmological constant satisfies
\ben
0&=&-(2+\epsilon)\lambda^*+\frac{1}{2}g^*(4\pi)^{-\epsilon/2}\left[
2(2+\epsilon)(3+\epsilon)\Phi_{1+\epsilon/2}^1(-2\lambda^*)
\right.\nonumber\\
&&\left.-8(2+\epsilon)\Phi_{1+\epsilon/2}^1(0)
-(2+\epsilon)(3+\epsilon)\epsilon\widetilde\Phi_{1+\epsilon/2}^1(-2\lambda^*)
\right]. \label{eq:FPL}
\een
If Eq. (\ref{eq:FPL}) is expanded in $\epsilon$, we have
\[
0=-2\lambda^*-2g^*\Phi_1^1(0)+O(\epsilon^2).
\]
Thus the UV NGFP for the cosmological constant is given by
\[
\lambda^*=-\frac{3}{38}\epsilon\Phi_1^1(0)+O(\epsilon^2).\label{eq:epsfpl}
\]
In contrast to the Newton constant, the cutoff function dependence appears
up to $O(\epsilon)$.

\section{Summary and discussion}
In this article we have discussed the existence of the UV NGFP in QG
by means of the ERGE.
The formulation of this article suggests that a UV NGFP exists in $2<d\leq 4$.
Thus it is possible that QG is an asymptotically safe theory and
renormalizable in $2<d\leq 4$.
The $\epsilon$-expansion of the $\beta$-function for the Newton constant
has a UV NGFP at $g^*=3\epsilon/38$ in $d=2+\epsilon$.
This coincides with the result of $2+\epsilon$ gravity theory in the harmonic
gauge up to $O(\epsilon)$. In addition it is independent of the shape of the
cutoff functions.
However, the cutoff function dependence will appear in $d>2+\epsilon$.
The $\epsilon$-expanded result of the UV NGFP for
the cosmological constant
has a cutoff function dependence up to $O(\epsilon)$. Thus there is the
possibility that the UV NGFP will disappear for other cutoff functions.
To resolve this problem it is necessary to perform
calculations for different sets of cutoff functions.\cite{Souma}

If the truncated ERGE is completely solved, these problems will
disappear. However it is not possible to solve it without any approximations.
For this reason, the functional space is approximated by the same space
of the Einstein-Hilbert action.
This approximation is the origin of the appearance of
the cutoff function dependence.
Thus it is expected that this dependence will be weakened by the extension
of the action functional space.
The first step to enlarge the functional space is to include the
so-called $R^2$-terms.\cite{Odintsov01} \ From another point of view,
the extension of the functional space is important.
This is because the RG flows that move toward the UV NGFP are governed by the
higher derivative terms. 
Hence to more accurately analyze the structure of phase space in the
UV regions,
treatment of these terms is necessary. 
However, this extension causes other problems.
If the functional space is approximated by the same space of $R^2$-gravity
theory, four $\beta$-functions will be derived, therefore,
four FPs will appear. 
However, one of these FPs will correspond to the UV NGFP derived in the
Einstein-Hilbert truncation.
In a perturbative method, such as the $\epsilon$-expansion, it is necessary
to calculate the higher loop corrections to get
reliable results. As is well known, even-loop calculations in the scalar
theory cause the UV NGFP in $d=4$. However, the scalar theory is believed
to be a trivial theory. Thus this UV NGFP is a fictitious FP.
These problems always appear in the perturbative method.
The truncation of the functional space in this article
is not a perturbative method.
In scalar theory,
the extension of the functional space does not cause the disappearance
of the NGFP in $3\leq d<4$.\cite{My} \ Thus it is expected that
the same situation will appear in quantum gravity.

As mentioned above, the gauge parameter $\alpha$ is
fixed to unity in this formulation.
However, it is possible to analyze it in other gauges.\cite{Odintsov02} \ The
$\beta$-function is not a physical quantity. Therefore, it depends on
the gauges.
Hence the UV NGFP which is derived from it has a gauge dependence.
However, this problem is not so serious if a UV NGFP exist for all
possible gauges.
This is because the important point is not the position of the UV NGFP
but its existence.
In the  Landau-De Witt gauge $\alpha=0$, 
it is recognized that the UV NGFP
remains in $d=4.$\cite{Odintsov01} \ However, to guarantee the existence
of the UV NGFP in $2<d\leq4$,
more accurate calculations must be done.\cite{Souma} \
In addition, there is the possibility that the physical quantities
derived from this formulation depend on the gauge. The corrections
from the constant gauge parameter are calculated in Ref.~\cite{Dou}, and
these calculations suggest that physical quantities such as the ratio of the
Newton constant and the cosmological constant are gauge independent if the
truncated ERGE is evaluated on-shell.

In addition to these problems, it is necessary to study the effectiveness of
this formulation.
As is well known, the physical quantities of QG have been exactly calculated
in $d=2$.
Hence it is important to calculate these quantities by means of the ERGE.
In this article, only the pure gravity theory is considered.
However, it is interesting to analyze the effects of the matter fields on
the structure of the phase space.
The ERGE including the
matter field has been derived in $d=4$.\cite{Dou} \ However, analysis for the
structure of the phase space does not exist.
\section*{Acknowledgements}
We would like to thank K.-I. Aoki,
H. Terao and T. Tabei for useful suggestions. 
Some of the numerical computations in this work were carried out 
at the Yukawa Institute Computer Facility.

\end{document}